\documentclass[report, twocolumn,nofootinbib,superscriptaddress]{revtex4} 
\usepackage{graphicx,longtable} 
\usepackage{dcolumn}
\usepackage{bm}
\usepackage{mathptmx}
\usepackage{amsmath}
\usepackage[pdftex]{hyperref}
\usepackage{wasysym}
\usepackage[sort&compress]{natbib}
\usepackage{float}
\usepackage{subfigure}

\newcommand{\beqy}{\begin{eqnarray}}
\newcommand{\eeqy}{\end{eqnarray}}
\newcommand{\bmlet}{\begin{subequations}}
\newcommand{\emlet}{\end{subequations}}
\newcounter{saveeqn}

\def\gsimeq{\,\,\raise0.14em\hbox{$>$}\kern-0.76em\lower0.28em\hbox  
{$\sim$}\,\,}  
\def\lsimeq{\,\,\raise0.14em\hbox{$<$}\kern-0.76em\lower0.28em\hbox  
{$\sim$}\,\,}  

\begin{document}

\title{First evidence of enhanced low-energy $\gamma$-ray strength from thermal neutron capture data}

\author{L.~Crespo~Campo}
\email{l.c.campo@fys.uio.no}
\affiliation{Department of Physics, University of Oslo, N-0316 Oslo, Norway}

\author{R.~B.~Firestone}
\affiliation{Department of Nuclear Engineering, University of California, Berkeley, CA 94720, USA}

\author{B.~A.~Brown}
\affiliation{National Superconducting Cyclotron Laboratory and Department of Physics and Astronomy, Michigan State University, East Lansing, Michigan 48824-1321, USA}

\author{M.~Guttormsen}
\affiliation{Department of Physics, University of Oslo, N-0316 Oslo, Norway}

\author{R.~Schwengner}
\affiliation{Institute of Radiation Physics, Helmholtz-Zentrum Dresden-Rossendorf, 01328 Dresden, Germany}

\date{\today}

\begin{abstract}

The $\gamma$-ray strength function, or average reduced $\gamma$-ray transition probability, is a fundamental input in the calculation of $(n,\gamma)$ cross sections used to simulate the nucleosynthesis of elements heavier than Fe. Since 2004, an enhanced probability of $\gamma$-decay with $\gamma$-ray energies below $\approx2 - 4$ MeV has been measured in reaction data for numerous nuclei. This has been observed as an increase in the $\gamma$-ray strength with decreasing $\gamma$-ray energy, often referred to as the low-energy enhancement or \textit{upbend} in the $\gamma$-ray strength. Nevertheless, the available data confirming this enhancement corresponded solely to charged-particle included reactions and no low-energy enhancement had yet been confirmed from neutron-induced reaction measurements. In this work,  we present the first evidence of low-energy $\gamma$-ray strength enhancement from neutron-capture reaction data. Gamma-ray spectra following thermal neutron capture on $^{58,60}$Ni have been used to determine the strength for primary and secondary $\gamma$-rays in $^{59,61}$Ni, showing an enhancement for $\gamma$-ray energies below $\approx 3$ MeV and $\approx 2$ MeV for $^{59,61}$Ni, respectively. For the first time, this enhancement is observed down to $\gamma$-ray energies of $\approx0.2$ MeV. Further, available spin-parity assignments have been used to obtain the multipolarity and electromagnetic character of these transitions, showing that this low-energy enhancement is dominated by $M1$ and $E2$ strength, with $E1$ strength also exceeding Standard Lorentzian Model predictions. Finally, large-basis shell-model calculations have been performed, also predicting a strong $M1$ enhancement at low $\gamma$-ray energies.

\end{abstract}

\maketitle

The investigation of nuclear $\gamma$-decay has been a fundamental tool for understanding single particle and collective nuclear structure phenomena~\cite{BohrVol1and2}. At sufficiently low excitation energies, discrete nuclear energy levels and their deexciting $\gamma$-ray transitions can be resolved using standard spectroscopy techniques. Level widths can be obtained and $\gamma$-ray transition probabilities can be determined. As the excitation energy of the nucleus increases, the number of quantum states becomes larger and average decay properties are used to describe the process of nuclear $\gamma$-decay. The $\gamma$-ray strength function ($\gamma$SF) is defined as the average reduced $\gamma$-ray transition probability in a given energy range; in other words, it describes how likely it is for a nucleus to absorb or emit $\gamma$-rays of a given energy~\cite{Bartholomew}. The $\gamma$SF is a necessary input in calculations of neutron-capture cross sections, which are of vital importance in fields such as nuclear reactor physics and nuclear astrophysics~\cite{daoutidis_goriely,63Ni_astrophysics,MyPaper64Ni}.

Above the neutron separation energy, $S_n$, the $\gamma$SF is dominated by the giant dipole resonance, and the accumulated $E1$ strength is generally well described with models such as the standard lorentzian (SLO) and generalized lorentzian (GLO)~\cite{Kopecky}. According to these models, the low-energy tail of the GDR still represents the largest contribution to the total strength at $\gamma$-ray energies below $S_n$ and presents a decrease with decreasing $\gamma$-ray energy. However, at low excitation energies $M1$ and $E2$ transitions are typically dominant, thus making the SLO and GLO descriptions inadequate in this region. 

In 2004, an increase in the dipole $\gamma$SF with decreasing $\gamma$-ray energy was first observed in $^{56,57}$Fe at $\gamma$-ray energies below $\approx2-4$ MeV. This low-energy enhancement or \textit{upbend} has been thoroughly investigated and seen as an excess of $\gamma$-ray strength when compared to the SLO and GLO predictions for $E1$ strength. Since its first observation in 2004, this low-energy enhancement has been measured in nuclides from Sc to Sm~\cite{Fe_Alex, 5051V_ld, Mo_RSF, 56Fe_Cecilie, Cd_strength_Cecilie, Sc, 43_Sc, 138_139_La_upbend, Sm_upbend} and confirmed with an independent method for $^{95}$Mo~\cite{wiedeking95Mo}. 

One of the greatest challenges in understanding this low-energy enhancement is the determination of the electromagnetic character and multipolarity of the associated $\gamma$ rays. For $^{60}$Ni there are strong indications of the $M1$ character of this upbend~\cite{60Ni_M1_upbend} and measurements for $^{56}$Fe suggest the enhancement is due to dipole transitions~\cite{56Fe_Cecilie}. However, the mechanisms underlying these enhancements are still unclear and different theoretical interpretations have been proposed. Shell-model calculations  for $^{94-96}$Mo and $^{90}$Zr suggest a strong low-energy enhancement of $M1$ transitions due to the recoupling of high-$j$ nucleon orbits~\cite{schwenger_upbend_M1_microcalc}. An $M1$ character of the low-energy enhancement has also been obtained from large-basis shell-model calculations on $^{56,57}$Fe~\cite{brown_upbend_M1_microcalc}, $^{60,64,68}$Fe~\cite{schwenger_upbend2017}, $^{73,74,80}$Ge~\cite{Therese737480Ge} and $^{89}$Y~\cite{CecilieY89}. Conversely, calculations based on the quasiparticle random phase approximation (QRPA) predict enhancement of $E1$ strength in $^{94,96,98}$Mo~\cite{litvinova_E1_upbend_94Mo} in agreement with the results from shell-model calculations for $^{44}$Sc~\cite{SiejaUpbendE1SM}. 


The experimental observation of this enhancement and the determination of the electromagnetic character and multipolarity of the associated $\gamma$-rays are of great importance. Not only can they give an insight into the mechanisms underlying the enhancement, but they can also impact the estimation of neutron-capture cross sections. If present in more neutron-rich, unstable nuclei, the low-energy enhancement may increase rapid-neutron capture  rates by up to two orders of magnitude~\cite{larsen_goriely_upbend_cross_section}.  Rapid-neutron capture has been considered responsible for creating $\approx 50\%$ of the nuclides in the solar system heavier than Fe~\cite{rprocessBurbridge, rprocessCameron} although the r-process sites had for long remained an open question. Recently, direct evidence of the r-process was experimentally observed, as both gravitational and electromagnetic waves were detected from a neutron-star merger event with the Advanced LIGO and Advanced Virgo detectors~\cite{LIGOneutronStarMerger}. With the low-energy enhancement highly affecting the results of $(n,\gamma)$ calculations, further investigation of this phenomenon is fundamental for our understanding of how heavy-elements are formed~\cite{larsen_goriely_upbend_cross_section,70Ni_upbend_MSU, rprocessCalculationsArnould, rprocessCalculationsSurman}. 
 

The experimental data leading to the observation of this enhancement has so far been provided only by charged-particle induced reactions such as those investigated in the Oslo method~\cite{gutt_unfolding, Gut87_first_generation, normalization_Schiller00, systematic_errors}. So far no evidence of this low-energy enhancement has been obtained from measurements of neutron-induced reactions. The analysis of two-step $\gamma$-cascades  following thermal neutron capture in $^{95}$Mo shows no evidence of such enhancement in the $\gamma$SF of $^{96}$Mo \cite{MilanNoUpbend96Moneutron}. Given the potential impact of the upbend in neutron-capture cross section, it is of vital importance to prove whether this phenomenon is present in $\gamma$-decay following neutron-induced reactions.

In this work we provide the first evidence of an enhanced low-energy $\gamma$-ray strength from neutron-capture measurements. To do so, thermal neutron capture data on $^{58,60}$Ni~\cite{Raman} have been used to estimate the $\gamma$-ray strength of $^{59,61}$Ni. From the primary and secondary $\gamma$ - ray energies and transition probabilities, the individual $\gamma$-ray transition strengths have been determined using the available level widths and lifetimes~\cite{ENSDF}. Further, tabulated spin-parity assignments have been used to obtain the multipolarity and electromagnetic character of these transitions.
Although each individual transition strength can fluctuate widely, the neutron capture decay schemes are nearly complete and several transition strengths of similar energies can be averaged for each multipolarity. 

As a result, the low-energy enhancement has been found in $^{59,61}$Ni and it can be ascribed primarily to $M1$ transitions.  Finally, the results have been compared to other experimental data for these nuclides and to the results of shell-model calculations.


The data analyzed in this work were obtained by Raman et al.~\cite{Raman}, where the experimental method is discussed in detail. As shown there, $\gamma$-decay spectra for $^{59,61}$Ni were measured following thermal neutron capture by the stable isotopes $^{58,60}$Ni. The experimental data were obtained by irradiating both enriched and natural Ni targets  in the thermal column of the internal target facility at the Los Alamos Omega West Reactor. In both cases, the targets were placed 1.5 m from the edge of the reactor core, receiving a thermal-neutron flux of $\sim6\cdot10^{11}$cm$^{-2}$s$^{-1}$. Gamma-ray spectra were obtained with a 30-cm$^3$ coaxial intrinsic Ge detector positioned inside a 20-cm-diameter by 30-cm-long NaI(Tl) annulus located at 6.3 m from the target. Energy calibrations were performed with a melamine (C$_3$H$_6$N$_6$) comparator, and the thermal neutron capture $\gamma$-ray cross sections were calibrated with a 100.0-mg CH$_2$ standard. As a result, a total of 414 primary and secondary $\gamma$-rays were placed in the level scheme of $^{59}$Ni and 240 $\gamma$-rays were placed for $^{61}$Ni, giving nearly complete level schemes for both isotopes.

If the decay scheme is complete and internal conversion can be neglected the total cross section of primary $\gamma$-rays, $\sum \sigma_{\gamma}^{\rm{primary}}$, equals within uncertainties the total cross section feeding the ground state, $\sum \sigma_{\gamma}^{\rm{GS}}$. For $^{59}$Ni $\gamma$-rays $\sum \sigma_{\gamma}^{\rm{primary}}=4.14\pm0.04$ b and $\sum \sigma_{\gamma}^{\rm{GS}}=4.11\pm0.05$ b, while for $^{61}$Ni $\sum \sigma_{\gamma}^{\rm{primary}}=2.308\pm$0.018 b and $\sum \sigma_{\gamma}^{\rm{GS}}$=2.39$\pm$0.03 b. Further, the observed unplaced $\gamma$-ray cross section is just 1.9$\%$ of the total for $^{59}$Ni and 1.6\% for $^{61}$Ni. 


The transition probabilities $P_{\gamma}$ were obtained as $P_{\gamma}=\frac{\sigma_{\gamma}}{\sigma_0}$ for primary $\gamma$-rays, where $\sigma_{\gamma}$ is the cross section for a given $\gamma$-ray transition and $\sigma_0$ is the total thermal neutron capture cross section, here $\sigma_0=4.13\pm0.05$ b and $\sigma_0=2.34\pm0.05$ b for $^{59}$Ni and $^{61}$Ni, respectively~\cite{Raman}. Secondary $\gamma$-ray transition probabilities  were obtained as $P_{\gamma}=\frac{\sigma_{\gamma} (1+\alpha)}{\sum_{i=1}^{} \sigma_{\gamma_i} (1+\alpha_i)}$ where the sum runs over all the $\gamma$-rays deexciting a given energy level and $\alpha$ is the internal conversion coefficient.

For each individual transition, we define the reduced $\gamma$-ray strength as $f_L=\frac{P_{\gamma}\Gamma_{\rm{tot}}}{E_{\gamma}^{2L+1}}$, where $L$ is the multipolarity of the transition, $\Gamma_{\rm{tot}}$ is the total radiative width of the initial level and the factor $P_{\gamma}\Gamma_{\rm{tot}}$ is the partial radiative width, $\Gamma_{\gamma}$~\cite{Bartholomew}. For primary transitions the total radiative width at $S_n$ is obtained from s-wave neutron resonance widths and $\Gamma_{\rm{tot}}=\left< \Gamma_{\gamma,0} \right>$, where $\left< \Gamma_{\gamma,0}\right>=2030\pm800$ meV and $\left< \Gamma_{\gamma,0} \right>=1120\pm200$ meV for $^{59}$Ni and $^{61}$Ni, respectively~\cite{RIPL3}. In the case of secondary transitions, the width of a given initial state was obtained as  $\Gamma_{\rm{tot}}= \frac{\hbar\cdot\ln2}{t_{1/2}}$, where $t_{1/2}$ is the half life of the state, obtained from ENSDF evaluations~\cite{ENSDF}. Since not all secondary levels had measured half-lives, only a subset of transitions from levels of known half-life are considered in this work.

\begin{figure}[bt]
	\includegraphics[clip,width=1.00\columnwidth]{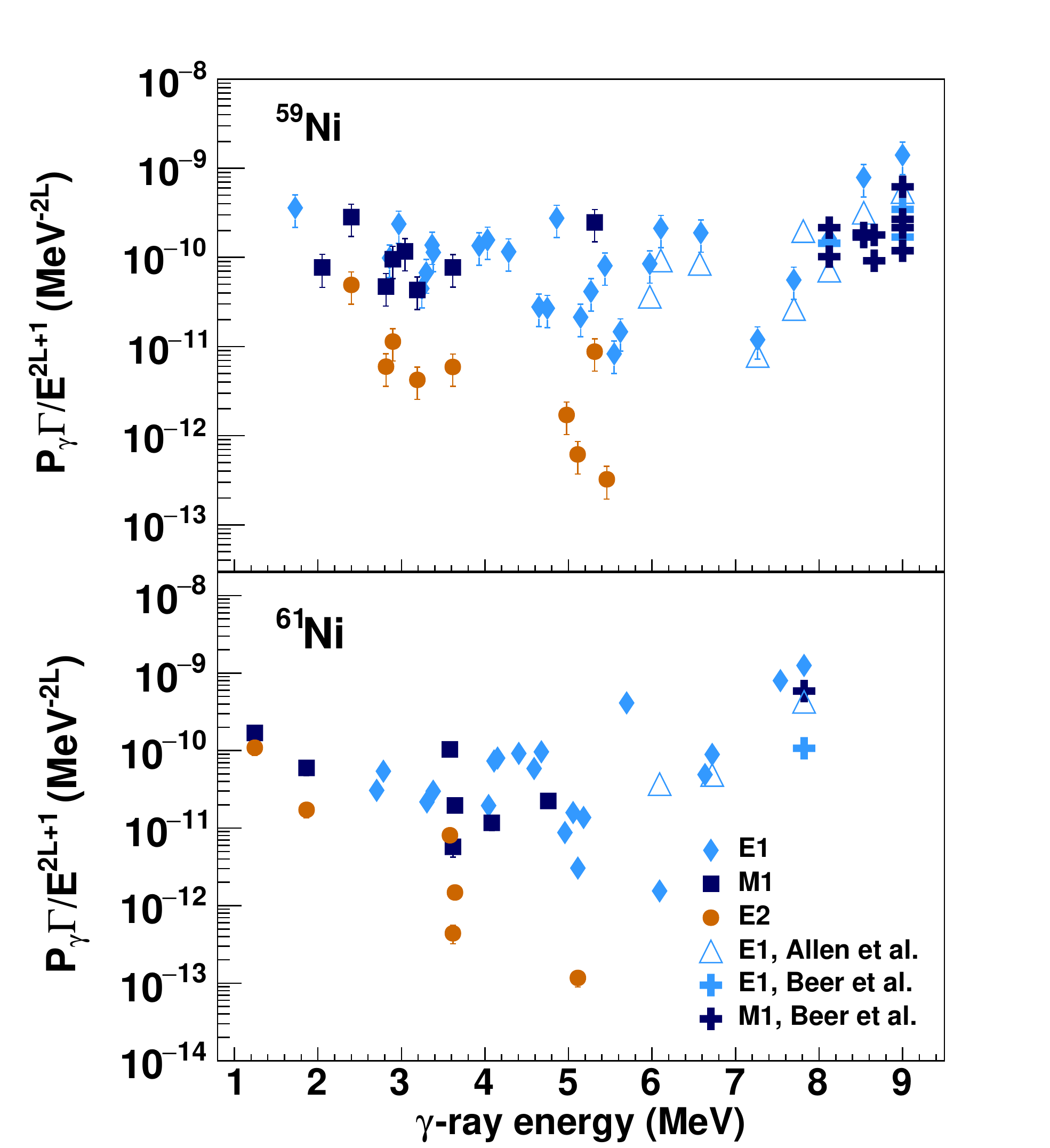}
	\caption{(Color online) Reduced strength for individual primary transitions in $^{59}$Ni (left) and $^{61}$Ni (right).}
	\label{fig:Primaries}
\end{figure}

\begin{figure}[bt]
	\includegraphics[clip,width=1\columnwidth]{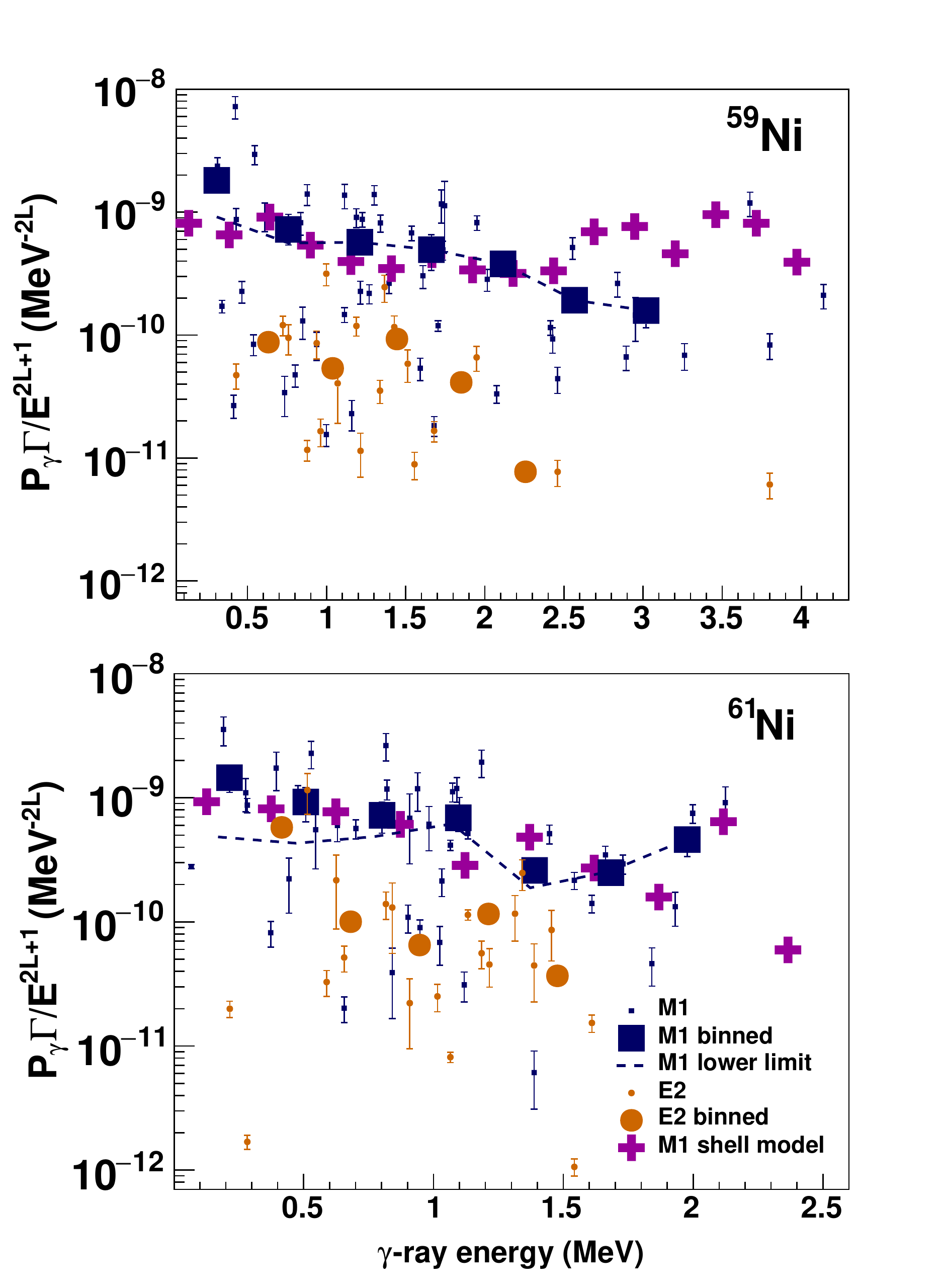}
	\caption{(Color online) Reduced strength for individual secondary transitions in $^{59}$Ni (left) and $^{61}$Ni (right). The results for $M1$ and $E2$ transitions have been averaged in bins of 450 keV and 300 keV width for $^{59,61}$Ni, respectively. In addition, the effect of weak, unobserved transitions has been estimated by assigning them a strength equal to the minimum strength observed per transition.}
	\label{fig:Secondaries}
\end{figure}

\begin{figure}[bt]

	\includegraphics[clip,width=1.0\columnwidth]{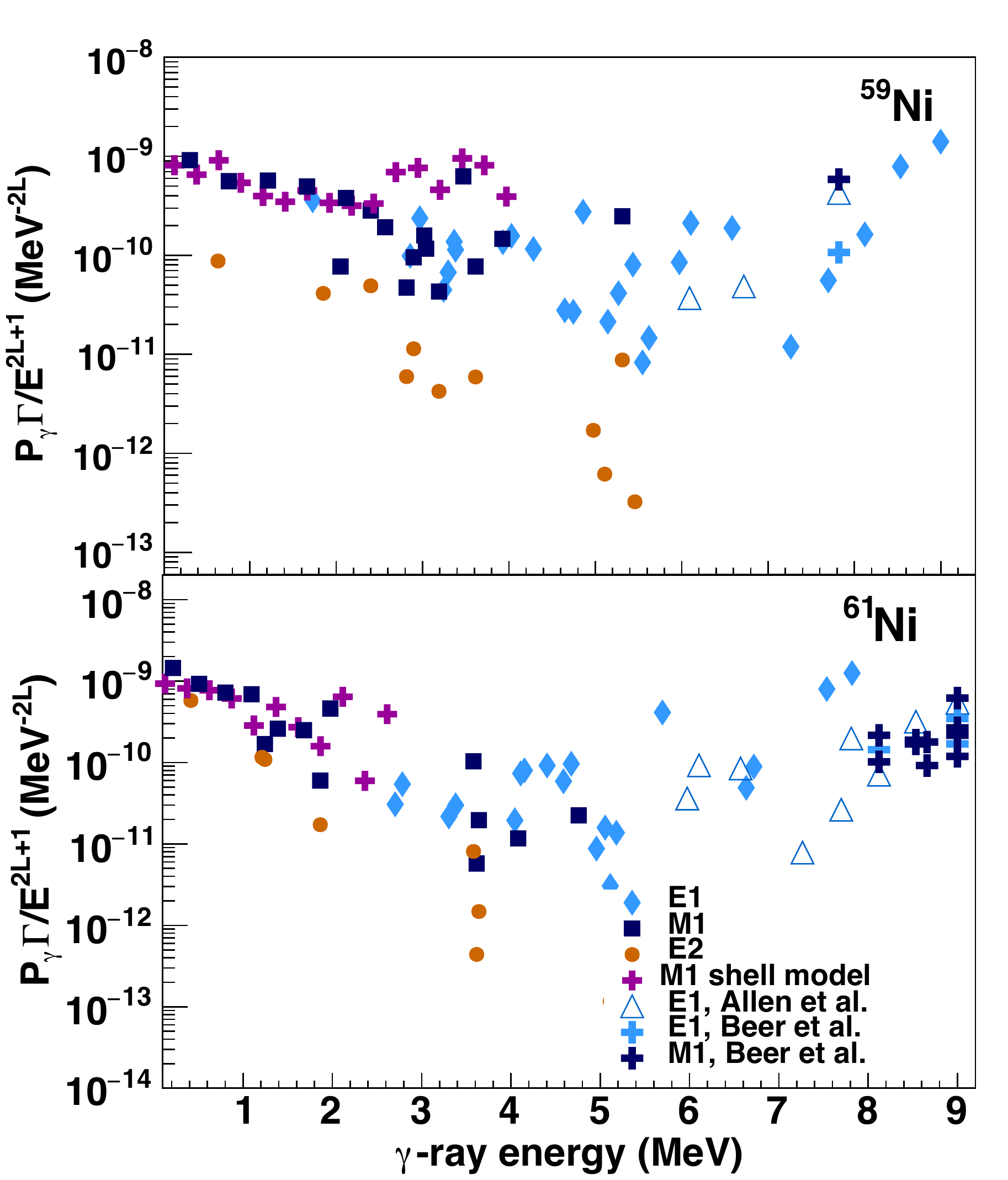}
	\caption{(Color online) Reduced strength for individual transitions in $^{59}$Ni (left) and $^{61}$Ni (right). Both primary and secondary transitions are shown. The observed low-energy enhancement of $M1$ transitions is compared to shell model calculations. Finally, additional resonances from Refs.~\cite{NiResAllen} and~\cite{NiResBeer} are included for $^{59}$Ni. }
	\label{fig:BothNiAll}
\end{figure}

Using spin-parity assignments for initial and final levels, as well as  mixing ratios~\cite{ENSDF}, the multipolarity ($L$) and electromagnetic character ($X$)  of the transitions were determined. The estimated transition strengths were then obtained for each $XL$, including results for primary transitions and secondary transitions from levels with known $t_{1/2}$. Note that, when no mixing ratios were available, the lowest mulitpolarity $XL$ was assumed to dominate. The resulting strengths for primary transitions are shown in Fig.~\ref{fig:Primaries}, while the strengths for secondary transitions are included in Fig.~\ref{fig:Secondaries}. In addition, a comparison of both primary and secondary transition strengths is shown in Fig.~\ref{fig:BothNiAll}. 

The quantity $f_L$ is proportional to the reduced $\gamma$-ray transition probability $B(XL)$, via \mbox{$B(XL)=(\hbar c)^{2L+1}\frac{L[(2L+1)!!]^2}{8\pi(L+1)}f_L$}, where \mbox{$\hbar c=1.973\cdot10^{-13}$ MeV$\cdot$m}. The results for $f_L$ presented in this work have been used to obtain the corresponding $B(XL)$ values in Weisskopf units, as shown in Appendix 1 to this manuscript.

As seen in Fig.~\ref{fig:Primaries}, both $^{59}$Ni and $^{61}$Ni exhibit comparable primary transition strengths, dominated by $E1$ transitions above $E_{\gamma}\approx4$ MeV. No positive parity levels are known below 3.2 MeV in $^{59}$Ni (where the level scheme is essentially complete) and, since its neutron separation energy is $S_n = 8.99$ MeV, the primary strength above 5.8 MeV is observed as nearly pure $E1$ for $^{59}$Ni. Similarly, no low-spin positive parity levels appear below 3 MeV for $^{61}$Ni and, since $S_n = 7.82$ MeV, nearly all primary $\gamma$-rays observed with $E_{\gamma}>4.8$ MeV are of $E1$ character. Above $E_{\gamma}\approx 5$ MeV,  the strengths fluctuate substantially for both nuclides, as they populate low excitation energies where fewer nuclear states are available. For instance, in $^{59}$Ni the transition strengths observed at $E_{\gamma}\approx 8.8-9$ MeV are up to two orders of magnitude higher that the transition strengths estimated at $E_{\gamma}\approx 7.3$ MeV. In contrast, smaller fluctuations are observed below $E_{\gamma}\approx 5$ MeV, where the $E1$ transition strengths remain approximately constant at $f_L\sim 10^{-10}$ MeV$^{-2}$. This evolution of the fluctuations with increasing number of nuclear states involved in the decay is expected if the measured partial radiative widths varied according to the Porter-Thomas distribution~\cite{PorterThomas}. For both $^{59,61}$Ni the observed primary dipole strength, obtained down to $E_{\gamma}=1.8, 1.3$ MeV, does not decrease with decreasing energy, in contrast to the SLO and GLO descriptions~\cite{Kopecky}.


 In addition to $E1$ transitions, several $M1$ transitions of comparable strength can be seen below $E_{\gamma}\approx5.5$ MeV in both $^{59,61}$Ni. Finally, some $E2$ transitions are present at $E_{\gamma}<6$ MeV, with increasing strength as the $\gamma$-ray energy decreases. The $E2$ transition strengths are up to $2-3$ orders of magnitude weaker than for dipole radiation at $E_{\gamma}\approx3-5$ MeV for both isotopes. However, $E2$ transitions with strength comparable to the dipole transitions are also seen, such as those measured at $E_{\gamma}\approx 2.4,5.3$ MeV in$^{59}$Ni and at $E_{\gamma}\approx 1.2, 3.6$ MeV in $^{61}$Ni. Note that the primary $\gamma$-rays populate levels below the continuum where the level energy separation is large enough to resolve all but the weakest transitions. Dipole primary $\gamma$-rays are observed to 19 of 20 possible final levels below 3 MeV in $^{59,61}$Ni suggesting that nearly all of the $\gamma$-ray strength is observed here. Finally, neutron resonance data from Refs.~\cite{NiResAllen, NiResBeer} have been used to estimate transition strengths for $^{59,61}$Ni. The results, included in Fig.~\ref{fig:Primaries}, are in agreement with the data analyzed in this work.

The reduced $\gamma$-ray strengths  for secondary transitions in $^{59,61}$Ni are shown in Fig.~\ref{fig:Secondaries}, all of which are of $M1$ or $E2$ character. Again, clear similarities are seen in both nuclei. The data for $M1$ are more scattered, with $f_L\sim10^{-11}-10^{-8}$ MeV$^{-2}$, which is indicative of a more complete set of transition strengths.  However, most of the $M1$ transitions are contained in the $f_L\sim10^{-10}-10^{-9}$ MeV$^{-2}$ range. The data for $M1$ secondary transitions has been binned (averaged) with bin widths of 450 and 300 keV for $^{59,61}$Ni. The results show a clear increase with decreasing $\gamma$-ray energy below $\approx 3$ and $2$ MeV in $^{59,61}$Ni, i.e., a low-energy enhancement or upbend in the measured $\gamma$-ray transition strengths. The secondary $M1$ strength near $E_{\gamma}\approx 0$ MeV for $^{59}$Ni is a factor of $\approx4$ larger than the strength at $E_{\gamma}\approx 2$ MeV, while for $^{61}$Ni the strength at $E_{\gamma}\approx 0$ MeV is 3 times larger than at $E_{\gamma}\approx 2$ MeV. 

The strengths of $E2$ transitions, averaged in 400 keV bins for both nuclei, are $\approx$ 2 orders of magnitude lower than for $M1$ transitions. As for $M1$ radiation, an increase in $E2$ strength with decreasing energy is seen. A similar trend is observed in the averaging of $E2$ transition strengths in $^{94,95}$Mo, obtained from shell model calculations~\cite{SchwengnerMoE2}.

To test the presence of enhanced $M1$ strength at low $\gamma$-ray energies, the impact of low-intensity transitions not observed in this work has been studied. In particular, a lower limit of the $\gamma$SF has been estimated as follows: based on the observed transitions, the expected weak transitions not observed in this work have been accounted for in the averaging, each one with an intensity equal to the minimum intensity measured. The resulting lower limit is shown as a dashed line in Fig.~\ref{fig:Secondaries}. For $^{59}$Ni, a clear increase of $M1$ strength with decreasing $\gamma$-ray energy is observed even in the estimated lower limit. In the case of $^{61}$Ni, no decrease is seen, although the average is substantially reduced with respect to the binned experimental data.

The experimental results presented in this manuscript have been compared to large basis shell-model calculations for the $M1$ strength. The shell-model calculations were carried out by means of the code NuShellX@MSU \cite{bro14Nu}. The CA48PN model space was used with the CA48MH1 Hamiltonian \cite{hjo95}, in which the single-particle energy of the $\nu 0g_{9/2}$ orbit was modified to fit better the Ni region. The model space included the $\pi(0f_{7/2}^{(8-t)}, 0f_{5/2}^t, 1p_{3/2}^t, 1p_{1/2}^t)$ proton orbits with $t$ = 0, 1, 2, and the 
$\nu(0f_{5/2}^{n5}, 1p_{3/2}^{p3}, 1p_{1/2}^{p1}, 0g_{9/2}^{g9})$ neutron orbits. The neutron occupation numbers were $n5$ = 0 to 3 for $^{59}$Ni and 1 to 5 for $^{61}$Ni, $p3$ = 0 to 3 for $^{59}$Ni and 0 to 4 for $^{61}$Ni, $p1$ = 0 to 2, and $g9$ = 0 to 2. In addition, effective $g$ factors of $g^{\rm eff}_s = 0.9  g^{\rm free}_s$ were applied. The calculations of $M1$ strengths included the lowest 40 states each with spins from $J_i^\pi, J_f^\pi$ = 1/2$^-$ to 21/2$^-$. The reduced transition strengths $B(M1)$ were calculated for all transitions from initial to final states with energies $E_f < E_i$ and spins $J_f = J_i, J_i \pm 1$. 
This resulted in more than 24500 $M1$ transitions. Since the experimental data correspond to primary transitions and secondary transitions from initial excitations energies below 4.2 MeV and 2.2 MeV for $^{59,61}$Ni, the shell-model calculations were gated to include decays from those excitation energies and below, as well as the primary $\gamma$-ray strengths. The transitions were then binned in $E_{\gamma}$, with bin widths of 250 keV and 240 keV for $^{59,61}$Ni respectively. The results, included in Figs.~\ref{fig:Secondaries} and~\ref{fig:BothNiAll}, show a good agreement with the experimental data. The upbend is successfully reproduced with the shell-model calculations, which also suggest the presence of strong $M1$ transitions above $E_{\gamma}\approx3$ MeV in the case of $^{59}$Ni.


\begin{figure}[bt]
	\includegraphics[clip,width=0.90\columnwidth]{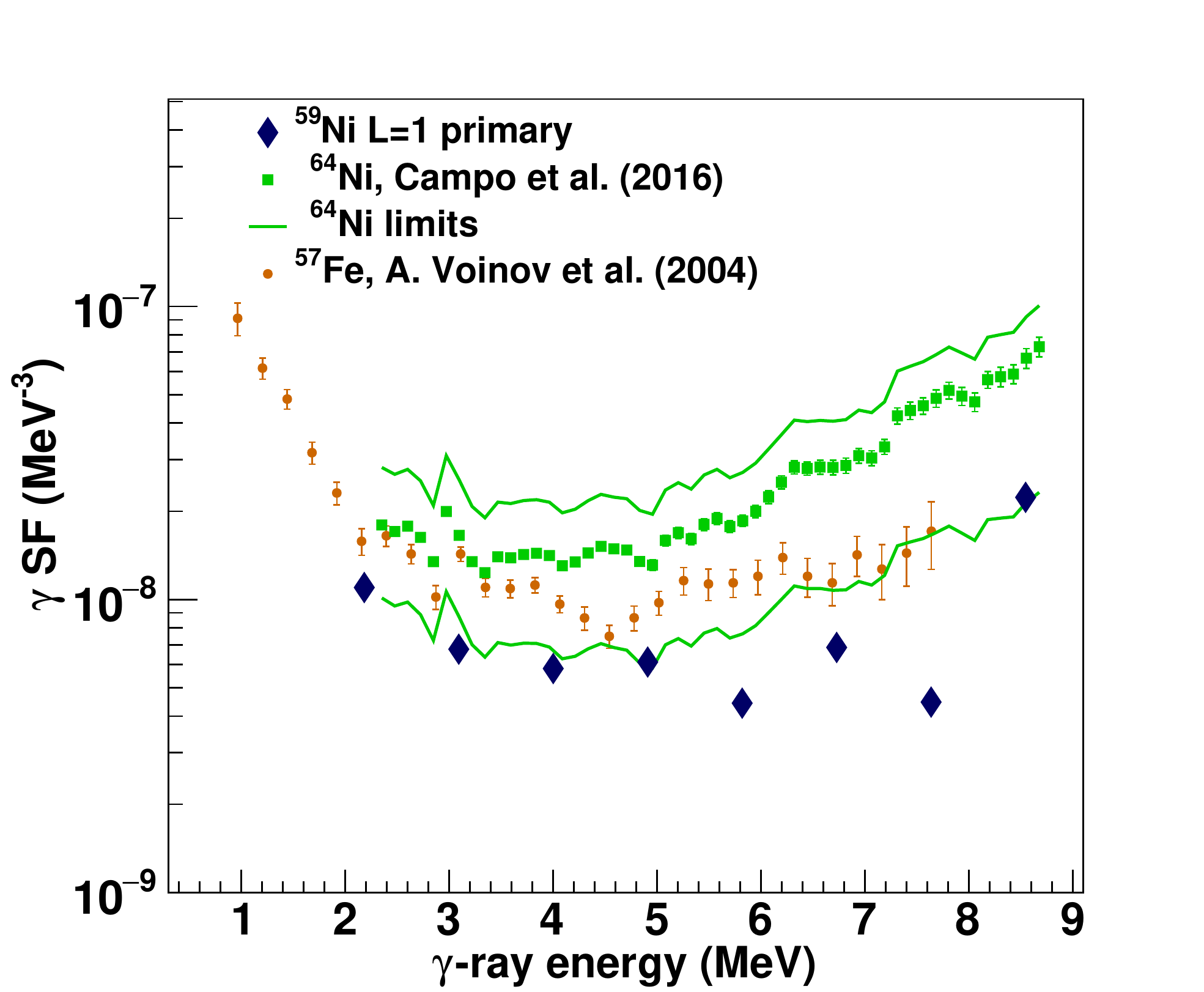}
	\caption{(Color online) Dipole $\gamma$SF from primary transitions in $^{59}$Ni compared to the experimental results for $^{57}$Fe~\cite{Fe_Alex} and $^{64}$Ni, obtained with the Oslo method~\cite{MyPaper64Ni}. }
	\label{fig:PrimariesD0GSF}
\end{figure}

Finally, it is important to note that the strength $f_L$ has units of MeV$^{-2}$, whereas the standard $\gamma$-ray strength function ($\gamma$SF), defined not per transition, but as an average per MeV, has units of MeV$^{-3}$. The reason behind this is that the $\gamma$SF includes a division by the nuclear level spacing for a given spin and parity, according to the original definition by Ref.~\cite{Bartholomew}. In this work, the $\gamma$SF in MeV$^{-3}$ has been estimated from primary transitions, since the level spacing at the neutron capture state $D_0$ is well defined. The estimates were obtained for $^{59}$Ni due to higher statistics and included in Fig.~\ref{fig:PrimariesD0GSF}. In addition, the results have been compared with the $\gamma$SFs of $^{57}$Fe~\cite{Fe_Alex} and $^{64}$Ni~\cite{MyPaper64Ni} obtained with the Oslo method, which present a clear low-energy enhancement below $E_{\gamma}\approx 3$ MeV. The primary dipole strength obtained for $^{59}$Ni is in reasonable agreement with the experimental data for $^{57}$Fe and $^{64}$Ni: it does not decrease with decreasing $\gamma$-ray energy (in contrast to the predictions of models such as the SLO) and present a slight increase below $E_{\gamma}\approx 4$ MeV.

In summary, we report on the first observation of an enhanced low-energy $\gamma$-ray strength from neutron-induced reaction measurements. Thermal neutron capture data on $^{58,60}$Ni have been used to estimate the strengths of $\gamma$-transitions in $^{59,61}$Ni. The results suggest that $E1$ strength does not decrease with decreasing $\gamma$-ray energies below $E_{\gamma}\approx4$ MeV, which exceeds the predictions of models such as the SLO, but is in agreement with the recent results of shell-model
calculations for $E1$ transitions~\cite{SiejaUpbendE1SM}. The observed $M1$ strengths of $^{59,61}$Ni present an increase with decreasing $\gamma$-ray energy below $\approx3, 2$ MeV. This low-energy enhancement is in agreement with the large basis shell-model calculations presented in this work.  Finally, an increase of $E2$ strength with decreasing $\gamma$-ray energy is also observed below $\approx 2$ MeV for $^{59,61}$Ni, as previously reproduced in shell model calculations of $^{94,95}$Mo~\cite{SchwengnerMoE2}.


We would like to thank A. C. Larsen and S. Siem for their valuable comments and suggestions. L. Crespo Campo acknowledges founding from the University of Oslo through the Kristine Bonnevie Scholarship and B. A. Brown acknowledges founding from NSF under grant PHY-1404442 RC103848. This work was performed in part under the auspices of the U.S. Department of Energy by the University of California, supported by the Director, Office of Science, Office of Basic Energy Sciences, of the U.S. Department of Energy at Lawrence Berkeley National Laboratory under Contract No. DE-AC02-05CH11231.

\end{document}


\title{Appendix 1
	
	Reduced $\gamma$-ray transition probabilities, $B(XL)$: experimental results and shell model calculations}

%

\author{L.~Crespo~Campo}
\affiliation{Department of Physics, University of Oslo, N-0316 Oslo, Norway}

\author{R.~B.~Firestone}
\affiliation{Department of Nuclear Engineering, University of California, Berkeley, CA 94720, USA}

\author{B.~A.~Brown}
\affiliation{National Superconducting Cyclotron Laboratory and Department of Physics and Astronomy, Michigan State University, East Lansing, Michigan 48824-1321, USA}

\author{M.~Guttormsen}
\affiliation{Department of Physics, University of Oslo, N-0316 Oslo, Norway}

\author{R.~Schwengner}
\affiliation{Institute of Radiation Physics, Helmholtz-Zentrum Dresden-Rossendorf, 01328 Dresden, Germany}

\date{\today}

\maketitle

The reduced transition strength presented in the manuscript, $f_L$, is proportional to the reduced $\gamma$-ray transition probability, $B(XL)$, via:
\begin{equation}
B(XL)=(\hbar c)^{2L+1}\frac{L[(2L+1)!!]^2}{8\pi(L+1)}f_L ,
\end{equation}
 where \mbox{$\hbar c=1.973\cdot10^{-13}$ MeV$\cdot$m}. Thus, $B(XL)$  can be easily obtained from $f_L$ as $B(XL)=C_{XL}f_L$, where $C_{XL}$ is a proportionality constant that depends on the multipolarity and electromagnetic character of the emitted radiation. With $f_{L}$ given in units of MeV$^{-2L}$, $B(EL)$ in e$^2$b$^L$, and $B(ML)$ in $\mu_n^2$b$^{L-1}$, the proportionality constants are: $C_{E1}=9.557\cdot10^{3}$ e$^2$b$\cdot$MeV$^2$, $C_{M1}=8.660\cdot10^{7}$ $\mu_n^2$$\cdot$MeV$^2$ and $C_{E2}=1.240\cdot10^{8}$ e$^2$b$^2$$\cdot$MeV$^4$.

  \begin{figure}[b]
  	\begin{center}
  	\includegraphics[clip,width=0.6\textwidth]{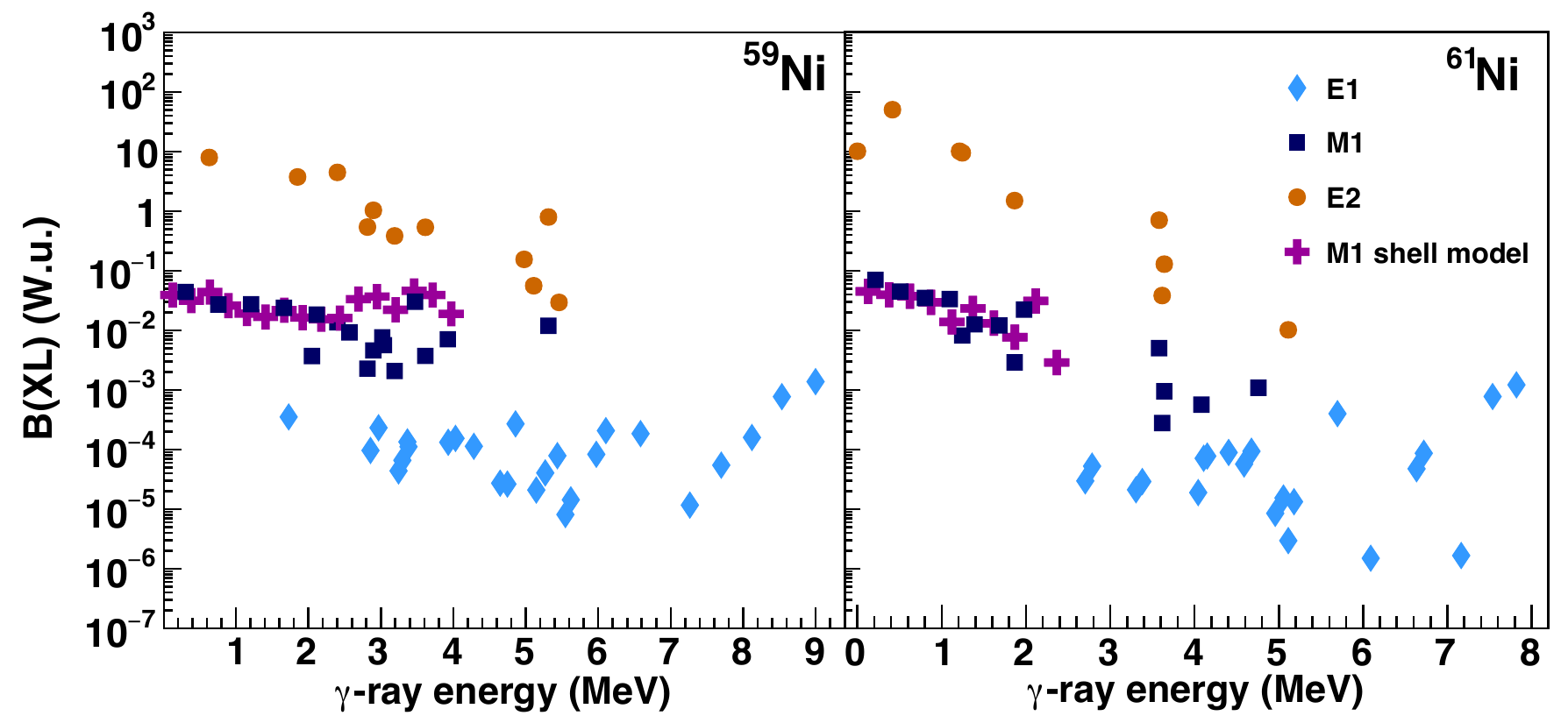}
  	\caption{Reduced $\gamma$-transition probabilities $B(XL)$ for $^{59}$Ni (left) and $^{61}$Ni (right), both for primary and secondary transitions. The values are shown in Weisskopf units (W. u.).}
  	\label{fig:AllWu}
  		\end{center}
  \end{figure}  

 In this work, $B(XL)$ values have been obtained from the transition strengths $f_L$ and expressed in Weisskopf units (W.u.). Note that 1 W.u.$=6.446\cdot10^{-4}A^{2/3}$ e$^2$b for electric dipole transitions, 1 W.u$=1.790$ $\mu_n^2$ for magnetic dipole transitions and \mbox{1 W.u.$=5.940\cdot10^{-6}A^{4/3}$ e$^2$b$^2$} for electric quadrupole transitions, with $A$ the mass number of the nucleus. The results are included in Fig.~\ref{fig:AllWu}, both for the experimental data as well as for the shell model calculations presented in this work.
 
Three different trends are seen for $E1$, $M1$ and $E2$ transitions. The observed $E1$ primary transitions have $B(E1)$ values on the order of $10^{-4}$ W.u. at all energies for both $^{59,61}$Ni. This is inconsistent with models such as the Standard Lorentzian (SLO), which predicts a rapid decline in strength at low $\gamma$-ray energies, generally below $E_{\gamma}\approx 2 - 4 $ MeV. Regarding $M1$ transitions, the enhancement below $E_{\gamma}\approx3$ MeV observed in the manuscript for secondary transition strengths ($f_L$), is also evident here. At $E_{\gamma}\approx 3 - 4$ MeV, $B(M1)$ values are as low as $\approx2\cdot10^{-3}$ W. u. and $3\cdot10^{-4}$ W. u. in $^{59, 61}$Ni, respectively. In contrast, $B(M1)$ values are $\sim0.1$ W. u. at $E_{\gamma}\approx100$ keV for both nuclides. Finally, some primary $E2$ transitions with $B(E2)\sim10^{-2}-10^{-1}$ W.u. are present above $E_{\gamma}>4$ MeV and $E_{\gamma}>3$ MeV in $^{59,61}$Ni. Below these energies, highly collective $E2$ transitions are observed: $B(E2)$ increases as $\gamma$-ray energy decreases, reaching  $\sim10^{1}-10^{2}$ W.u. at $E_{\gamma}\approx100$ keV.